\documentclass[12pt,a4paper]{article}
\usepackage{graphicx,amssymb,amsmath,color,scalefnt}
\usepackage{placeins}
\usepackage{multirow, makecell}
\usepackage{chngpage}
\usepackage{geometry}
\usepackage[dvipsnames]{xcolor}
\usepackage{rotating}
\usepackage[normalem]{ulem}
\usepackage{jheppub}
\notoc
\usepackage{scalerel}
\usepackage{makecell}
\usepackage{appendix}
\usepackage{stackengine}
\usepackage{booktabs}
\usepackage{comment}
\usepackage{subcaption}
\renewcommand{\thefootnote}{\fnsymbol{footnote}}

\makeatletter\g@addto@macro\bfseries{\boldmath}\makeatother

\DeclareMathAlphabet{\mathsfit}{\encodingdefault}{\sfdefault}{m}{sl}

\newcommand{\mailto}[1]{\href{mailto:#1}{#1}}

\begin{document}

\title{Sensitivity Analysis of the Top-Quark Sector}

\author[a]{Fernando Cornet-Gomez\footnote[1]{\mailto{fernando.cornet@uco.es}},}
\author[b]{V\'ictor Miralles\footnote[2]{\mailto{victor.miralles@ua.es}},}%
\author[c]{Marcos Miralles L\'opez\footnote[3]{\mailto{marcos.miralles.lopez@cern.ch}},}%
\author[d]{Mar\'ia Moreno Ll\'acer\footnote[4]{\mailto{maria.moreno@ific.uv.es}},}%
\author[d]{Marcel Vos\footnote[5]{\mailto{marcel.vos@cern.ch}}\,}

\affiliation[a]{Departamento de F\'isica, Universidad de C\'ordoba, Campus Universitario de Rabanales, Ctra. N-IV Km. 396, E-14071 C\'ordoba, Spain}
\affiliation[b]{Departament de F\'isica, Universitat d'Alacant,
Campus de Sant Vicent del Raspeig, E-03690 \mbox{Alacant}, Spain}
\affiliation[c]{School of Physics and Astronomy, University of Glasgow, Glasgow G12 8QQ, Scotland, United Kingdom}
\affiliation[d]{IFIC, Universitat de Val\`encia and CSIC, Calle Catedrático José Beltrén 2, E-46980 Paterna, Spain}

\def\TeV{\ifmmode {\mathrm{Te\kern -0.1em V}}\else
                   \textrm{Te\kern -0.1em V}\fi\,}%
\def\GeV{\ifmmode {\mathrm{Ge\kern -0.1em V}}\else
                   \textrm{Ge\kern -0.1em V}\fi\,}%
\def\MeV{\ifmmode {\mathrm{Me\kern -0.1em V}}\else
                   \textrm{Me\kern -0.1em V}\fi\,}%
\def\keV{\ifmmode {\mathrm{ke\kern -0.1em V}}\else
                   \textrm{ke\kern -0.1em V}\fi\,}%
\def\eV{\ifmmode  {\mathrm{e\kern -0.1em V}}\else
                   \textrm{e\kern -0.1em V}\fi\,}%
\let\tev=\TeV
\let\gev=\GeV
\let\mev=\MeV
\let\kev=\keV
\let\ev=\eV

\def\iab{\mbox{ab$^{-1}$}}
\def\ifb{\mbox{fb$^{-1}$}}
\def\ipb{\mbox{pb$^{-1}$}}
\def\inb{\mbox{nb$^{-1}$}}

\definecolor{Purple}{rgb}{0.6, 0.4, 0.8}
\newcommand{\vm}[1]{{ \color{orange} #1}}
\newcommand{\vmcom}[1]{{ \bf \color{Purple} #1}}
\newcommand{\com}[1]{{ \color{red} #1}}
\newcommand{\myComment}[1]{}
\newcommand{\mv}[1]{\textcolor{green!70!black}{#1}}
\newcommand{\fc}[1]{\textcolor{blue}{FCG: #1}}

\newcommand{\HEPfit}{\texttt{HEPfit }}
\newcommand{\ttbar}{{\ensuremath{t\bar{t}}}}

\renewcommand{\thefootnote}{\arabic{footnote}}

\abstract{
 We study the sensitivity of current and future collider observables to top-quark SMEFT operators through a one-operator-at-a-time analysis. Using data from the Tevatron, LEP, and LHC Run~2, as well as projections for the HL-LHC and future lepton colliders, we identify the measurements that provide the strongest individual constraints. This approach clarifies the role of specific observables in the top-quark SMEFT program and highlights the significant improvement in sensitivity expected at future facilities.
}

\maketitle

\section{Introduction}
The Standard Model Effective Field Theory (SMEFT) has become the primary framework for interpreting top-quark measurements at the Large Hadron Collider (LHC) and planning for future collider programs. By parameterizing potential new physics (NP) through higher-dimensional operators, the SMEFT allows for a model-independent quantification of the agreement between data and theory. 

While recent global fits~\cite{Cornet-Gomez:2025jot,deBlas:2025xhe} provide a comprehensive view of the current constraints on the top-quark sector \cite{Durieux:2019rbz,Miralles:2021dyw}, they often mask the specific sensitivity of individual observables due to the large number of degrees of freedom and the presence of ``blind directions'' in the parameter space. In a global analysis, correlations between Wilson coefficients (WCs) can significantly weaken the marginalized bounds, potentially obscuring the underlying precision of the experimental measurements. Indeed, in the case in which the NP does not follow the precise correlations pattern obtained in the global fits, the marginalized bounds would underestimate the constraints. 

In this report, we shift the focus from the global marginalized result to a detailed technical assessment of individual constraints. We present the 95\% probability intervals for a set of 29 Wilson coefficients, obtained by varying one operator at a time while fixing all others to their Standard Model values. This approach serves two primary purposes. First it identifies the ``golden channels'' and specific differential distributions that provide the most stringent constraints on particular operator classes. And second, it provides a baseline for experimentalists to evaluate the impact of new measurements.

We include current data from the Tevatron, LEP, and LHC Run 2, and provide projections for the High-Luminosity LHC (HL-LHC) and future lepton colliders, including $e^+e^-$ Higgs factories and the muon collider. By isolating the impact of individual observables, we demonstrate how future facilities could reach sensitivities for four-fermion operators as low as $\mathcal{O}(10^{-4})$ TeV$^{-2}$.

\section{Theoretical Framework and Methodology}
\label{sec:methodology}

The analysis is performed within the SMEFT framework, where the effective Lagrangian is expanded as:
\begin{equation}
\mathcal{L}_\text{eff} = \mathcal{L}_\text{SM} + \frac{1}{\Lambda^2} \sum_i C_i O_i + \mathcal{O}\left(\Lambda^{-4} \right) .
\end{equation}
Following the decoupling theorem~\cite{Appelquist:1974tg}, we focus on dimension-six operators built from SM fields, assuming $CP$ conservation\footnote{Some of the $CP$ violating interactions have also been studied in Ref.~\cite{Miralles:2024huv}.} and a flavor symmetry $U(2)^5$ that distinguishes the third generation~\cite{Faroughy:2020ina}. We employ the Warsaw basis~\cite{Grzadkowski:2010es} and follow the prescription of the LHC top-quark Working Group~\cite{AguilarSaavedra:2018nen} for the linear combinations of Wilson coefficients (WCs)\footnote{The full list of the WC considered can be found in Ref.~\cite{Cornet-Gomez:2025jot}}.

\subsection{Current Observables and Experimental Input}
\label{sec:current_obs}

The experimental foundation of this work is the dataset from the Tevatron, LEP, and LHC Run 2. This includes a wide array of top-quark processes: inclusive and differential $t\bar{t}$ production, single-top production in all channels ($t$, $s$, and $tW$), and associated production with gauge bosons ($t\bar{t}Z$, $t\bar{t}\gamma$, $t\bar{t}W$). 

As detailed in Ref.~\cite{Cornet-Gomez:2025jot}, our analysis also incorporated measurements of the quantum entanglement in the $t\bar{t}$ pair production at threshold~\cite{ATLAS:2023fsd,CMS:2024pts} and in the boosted regime~\cite{CMS:2024zkc}.
Although these measurements were part of the global baseline, their individual impact on the specific operators was already presented in ref.~\cite{Cornet-Gomez:2025jot} so it will not be discussed in this report.

\subsection{Future Projections}
\label{sec:projections}

To provide a roadmap for the top-quark sector, we include projections for the High-Luminosity LHC (HL-LHC)
, as well as future lepton collider. These include $e^+e^-$ Higgs factories (ILC and FCC-ee) and a high-energy muon collider operating at the  3-30~TeV range. The detailed simulation parameters for these future scenarios are provided in Ref.~\cite{Cornet-Gomez:2025jot}.

\subsection{Simulation and Fit Procedure}

In this technical report we made public the systematic mapping of individual WC sensitivities for each observable. While the global fits in Ref.~\cite{Cornet-Gomez:2025jot} provide the most statistically rigorous bounds by allowing all operators to vary simultaneously, they inevitably introduce correlations that can mask the inherent precision of a measurement. 

In the following section, we present the limits obtained including a single observable and by varying a single WC at a time. This sensitivity analysis allows for the identification of the most powerful observables for each operator, providing a clear diagnostic tool for understanding the current and future landscape of the top-quark SMEFT fit.

In general the physical observables contain a linear dependence on the WCs coming from the interference of the pure dimensión-six operators with the SM and a quadratic contribution from squaring dimension-six operators. 
In this report, we strictly adopt the linear truncation, including only terms proportional to $\Lambda^{-2}$. This approach ensures a conservative and theoretically consistent treatment, as a full $\mathcal{O}(\Lambda^{-4})$ analysis would require the inclusion of dimension-eight operator interferences~\cite{Brivio:2022pyi}.

The linear coefficients $X^{\rm{int}}_i$ are obtained using \texttt{MadGraph5\_aMC@NLO}~\cite{Alwall:2014hca} with the \texttt{SMEFTsim}~\cite{Brivio:2020onw} and \texttt{SMEFT@NLO}~\cite{Degrande:2020evl} models. Statistical inference is carried out via a Bayesian global fit implemented in the \texttt{HEPfit} \cite{DeBlas:2019ehy}  package.

\section{Results: Individual Sensitivity Breakdown}
\label{sec:results}

\begin{figure}[h!]
    \centering
    \includegraphics[width=0.9\linewidth]{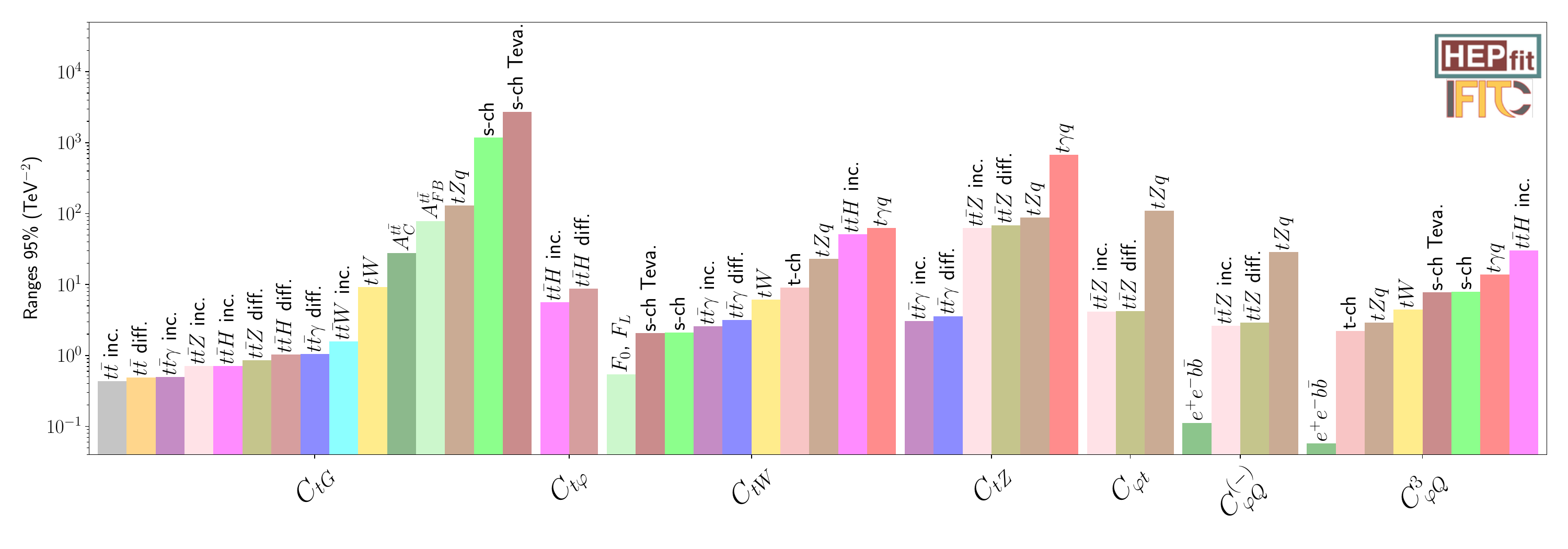}\\
    \includegraphics[width=0.9\linewidth]{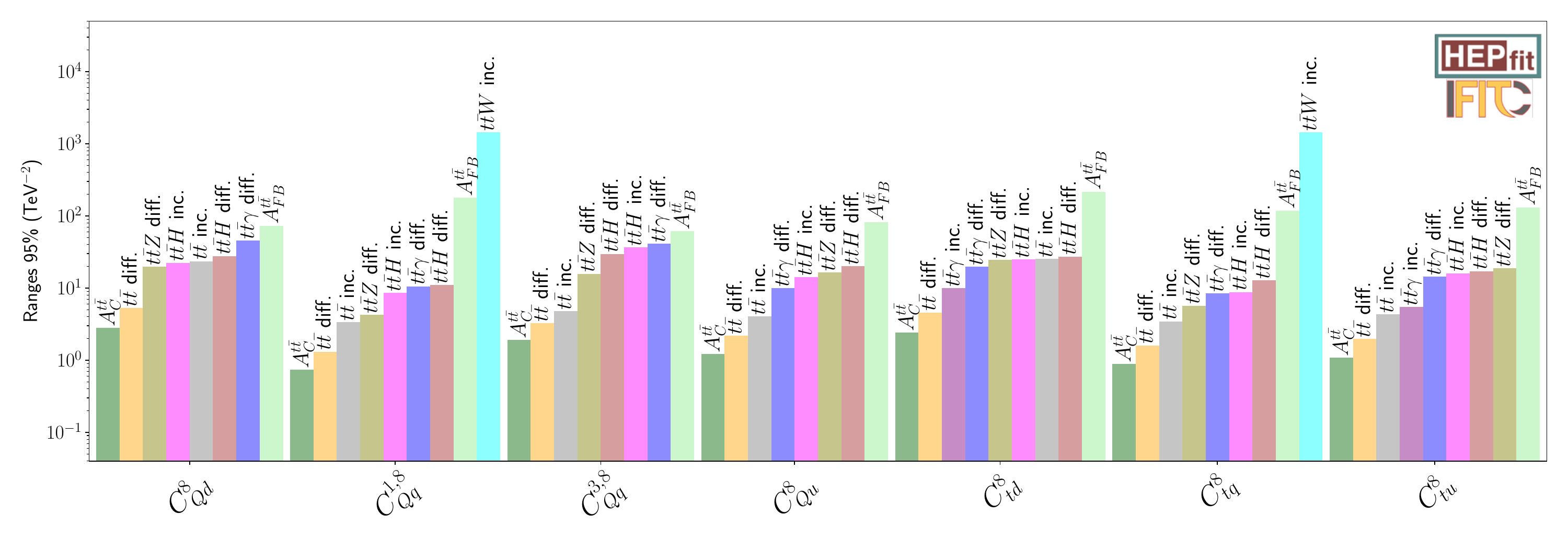}
    \caption{Comparison of the individual 95\% probability bounds for the 2-quark (top) and 4-quark operators octets (bottom), derived from the different measurements at current colliders (LHC, LEP and Tevatron).  The individual bounds are obtained from fits of a single operator coefficient to a single measurement. Similar results are obtained for the singlet coefficients. }
    \label{fig:current}
\end{figure}

In this section, we break down the 95\% probability intervals according to the experimental facility providing the constraint. By isolating the impact of each observable for the different future colliders, we demonstrate how the top-quark SMEFT landscape could evolve depending on which machine the community decides to build.

\subsection{Current Constraints and HL-LHC Projections}
As discussed, current experimental bounds relies on the Tevatron, LEP, and LHC Run 2 datasets. The individual sensitivities can be found on figure \ref{fig:current}. The four-quark sector is predominantly constrained by the $t\bar{t}$ charge asymmetry and the differential $t\bar{t}$ distribution at the LHC. The landscape at the HL-LHC (see figure \ref{fig:HL_LHC})
is similar but with an overall improvement of a factor of roughly 3. For some of the 2-quark operators, like $C_{\varphi Q}^{(-)}$ and $C_{\varphi Q}^{(3)}$, the most constraining observable comes from LEP. Although HL-LHC will improve LHC bounds, it will not be able to surpass the one that comes from the $e^+e^-\to b\bar{b}$ observable. The HL-LHC is also expected to give bounds to 2-quark 2-lepton operators as $C_{lt}, C_{eq}, C_{lq}^{(-)}$ and $C_{et}$

\begin{figure}
    \centering
    \includegraphics[width=0.9\linewidth]{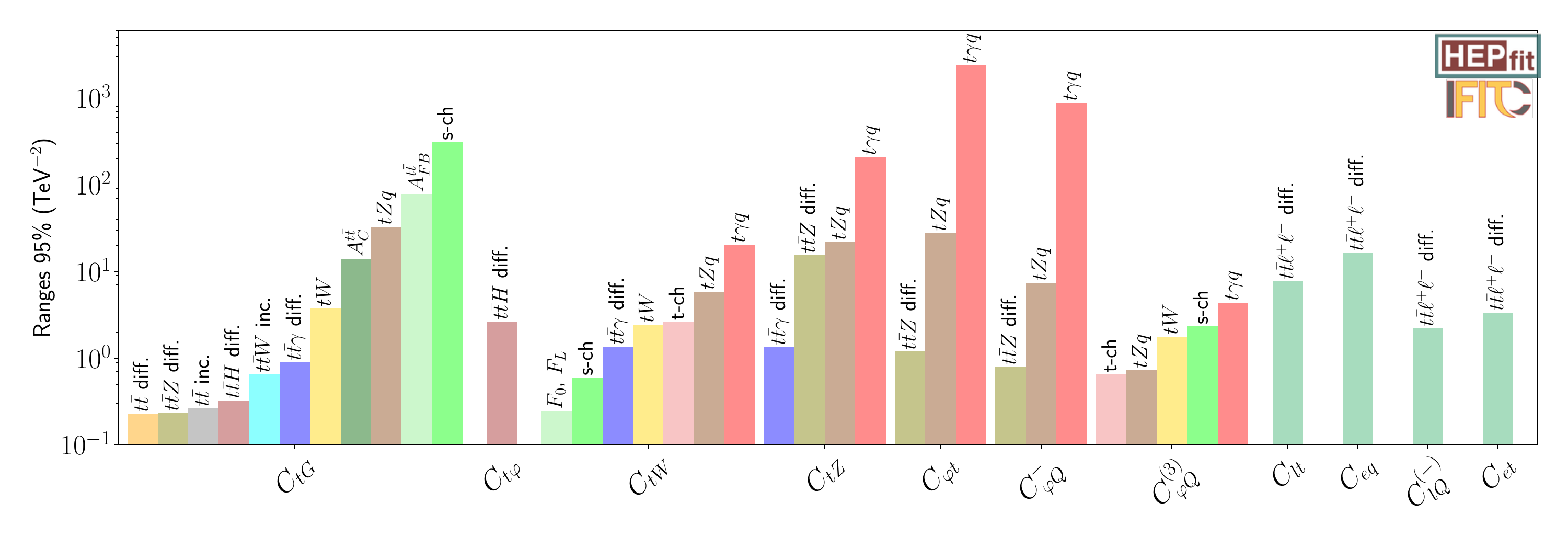}\\
    \includegraphics[width=0.9\linewidth]{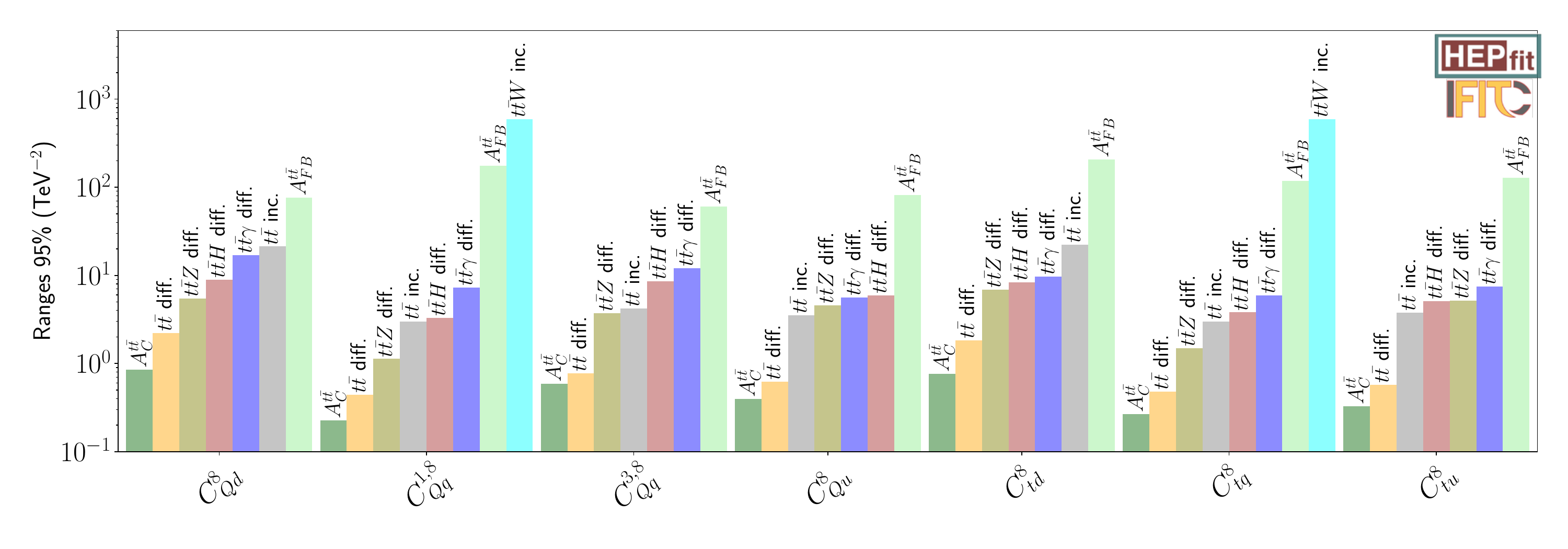}
    \caption{Comparison of the individual 95\% probability bounds for the 2-quark (top) and 4-quark operators octets (bottom), derived from the different measurements at the HL-LHC.  The individual bounds are obtained from fits of a single operator coefficient to a single measurement. Similar results are obtained for the singlet coefficients.}
    \label{fig:HL_LHC}
\end{figure}

\subsection{Future Lepton and Muon Colliders}
Future facilities offer a clean environment to reach sensitivities inaccessible to hadronic machines. The $e^+e^-$ Higgs Factories (ILC/FCC-ee) are essential for constraining the two-quark-two-lepton sector, as shown in figures \ref{fig:ILC} and  \ref{fig:FCC}. Our results show that they can achieve enough sensitivities to constrain some operators up to $\mathcal{O}(10^{-4})$~TeV$^{-2}$ like $C_{lu}$ and $C_{eq}$.

There have also been interest in the community for a possible future muon collider. The main advantage is the possibility of operating at much higher energies. Indeed, a muon collider operating in the 3--30~TeV range represents the ultimate frontier for vertex corrections (${C}_{tG}$, $C_{\varphi Q}^{(-)}$) and dipole terms as can be shown in figure \ref{fig:MuC}.


\begin{figure}
    \centering

    \begin{subfigure}{0.95\linewidth}
        \centering
        \includegraphics[width=\linewidth]{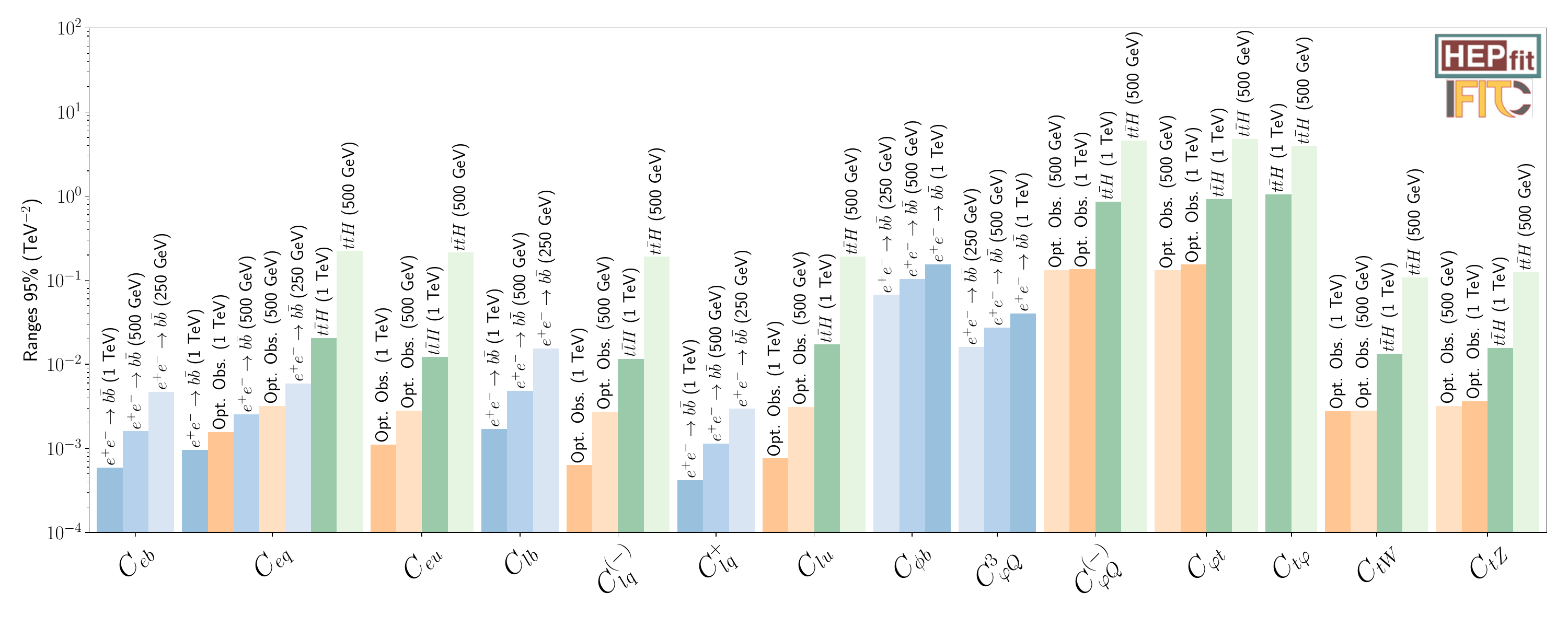}
        \caption{ILC}
        \label{fig:ILC}
    \end{subfigure}
    
    \vspace{0mm}
    
    \begin{subfigure}{0.95\linewidth}
        \centering
        \includegraphics[width=\linewidth]{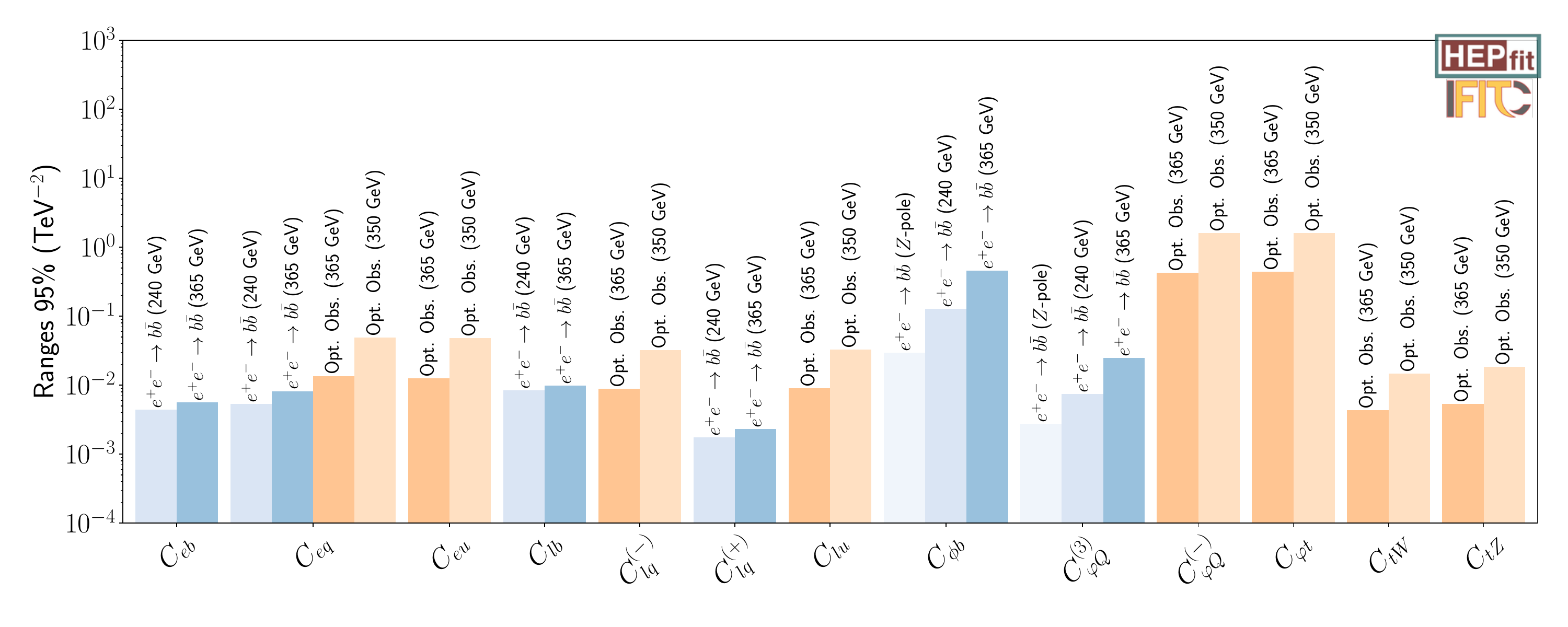}
        \caption{FCC-ee}
        \label{fig:FCC}
    \end{subfigure}
    
    \vspace{0mm}
    
    \begin{subfigure}{0.95\linewidth}
        \centering
        \includegraphics[width=\linewidth]{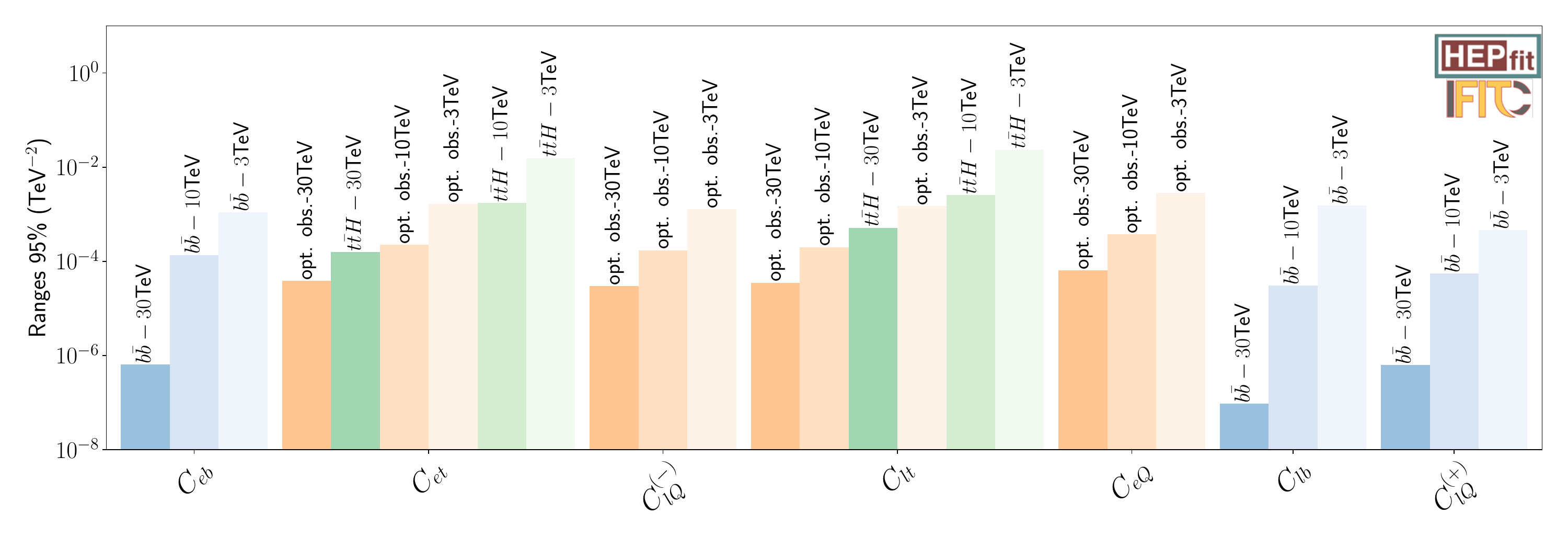}
        \caption{Muon Collider}
        \label{fig:MuC}
    \end{subfigure}

    \caption{Comparison of the individual 95\% probability bounds derived from different measurements at future lepton colliders: (a) ILC, (b) FCC-ee, and (c) Muon Collider. The individual bounds are obtained from fits of a single operator coefficient to a single measurement. For the Muon Collider case, the 2-fermion coefficients can be found in Ref.~\cite{Cornet-Gomez:2025jot}.}
    \label{fig:future_colliders}
\end{figure}

\FloatBarrier

\section*{Acknowledgments}
FCG and VM are supported by the Spanish \textit{Ministerio de Ciencia, Innovaci\'on y Universidades} through a Beatriz Galindo Junior grants BG23/00061 and BG24/00038, respectively. FCG is also suported by the 
The work of MML is supported by the Science and Technology Facilities Council [grant number ST/X005941/1]. 
MMLL acknowledges the support received from the Spanish Ram\'on y Cajal programme (RYC2019-028510-I). Her work is also supported by the projects ASFAE/2022/010 and CIPROM/2022/70 (Generalitat Valenciana), and PID2021-124912NB-I00 (Spanish Ministry MICIN). 
The work of MV is supported by the Spanish ministry for science under grant number PID2021-122134NB-C21, by the Generalitat Valenciana under PROMETEO grant CIPROM/2021-073, and by CSIC under grant ILINKB20065. The group members of IFIC in Valencia received support from the Severo Ochoa excellence programme.

\bibliographystyle{apsrev4-1_title}
\bibliography{top.bib}

\end{document}